\documentclass[11pt,a4paper]{article}
\usepackage{graphicx}
\usepackage{url}
\usepackage{amssymb}
\usepackage[centertags]{amsmath}
\usepackage{multirow}
\usepackage{subfigure}
\usepackage[affil-it]{authblk}
\usepackage{cite}
\usepackage{notoccite}
\usepackage{lineno}
\usepackage{tocbibind}
\usepackage[colorlinks,linkcolor=blue,anchorcolor=blue,citecolor=blue,urlcolor=blue]{hyperref}

\newcommand{\nuebar}{$\overline{\nu}_{e}$}

\begin{document}

\title{Prospects of detecting the reactor \nuebar-Ar coherent elastic scattering with a low threshold dual-phase argon time projection chamber at Taishan}

\author[1,2]{Yu-Ting Wei}
\author[1]{Meng-Yun Guan}
\author[1]{Jin-Chang Liu\footnote{liujinc@ihep.ac.cn}}
\author[1]{Ze-Yuan Yu\footnote{yuzy@ihep.ac.cn}}
\author[1,2]{Chang-Gen Yang}
\author[1]{Cong Guo}
\author[1,2]{Wei-Xing Xiong}
\author[1,2]{You-Yu Gan}
\author[1,2]{Qin Zhao}
\author[1,3]{Jia-Jun Li}

\affil[1]{Institute of High Energy Physics, CAS, Beijing 100049, China}
\affil[2]{University of Chinese Academy of Sciences, Beijing 100049, China}
\affil[3]{North China Electric Power University, Beijing 100096,China}

\maketitle

\begin{abstract}
 We propose to measure the coherent elastic neutrino-nucleus scattering~(CE$\nu$NS) using a dual-phase liquid argon time projection chamber~(TPC) with 200~kg fiducial mass.
 The detector is expected to be adjacent to the JUNO-TAO experiment and to be about 35~m from a reactor core with 4.6~GW thermal power at Taishan.
 The antineutrino flux is approximately 6$\times10^{12}$~cm$^{-1}$s$^{-1}$ at this location, leading to more than 11,000 coherent scattering events per day in the fiducial mass.
 However, the nuclear recoil energies concentrate in the sub-keV region, corresponding to less than ten ionization electrons in the liquid argon.
 The detection of several ionization electrons can be achieved in the dual-phase TPC due to the large amplification in the gas region.
 With a feasible detection threshold of four ionization electrons, the signal rate is 955 per day.
 The detector is designed to be shielded well from cosmogenic backgrounds and ambient radioactivities to reach a 16\% background-to-signal ratio in the energy region of interest.
 With the large CE$\nu$NS sample, the expected sensitivity of measuring the weak mixing angle $\sin^{2}\theta_{W}$, and of limiting the neutrino magnetic moment are discussed.
 In addition, a synergy between the reactor antineutrino CE$\nu$NS experiment and the dark matter experiment is foreseen.

\end{abstract}

keywords: {coherent elastic neutrino-nucleus scatting, dual-phase argon TPC, reactor antineutrinos}

\section{Introduction}

There has been over forty years since D.~Freedman predicted the coherent elastic neutrino-nucleus scattering process~(CE$\nu$NS)~\cite{PhysRevD.9.1389}.
In the weak neutral current, the $\nu+A \rightarrow \nu+A$ elastic scattering process should have a sharp coherent forward peak for $qR\ll1$.
Here, $q$ is the absolute momentum transfer from the neutrino to the nucleus, and $R$ is the weak nuclear radius.
The CE$\nu$NS is dominated interaction for neutrinos with energies less than $\sim$100~MeV.
It has been firstly observed at a $6.7\sigma$ confidence level with a CsI[Na] detector in the COHERENT experiment~\cite{Akimov:2017ade}.
It used pulsed neutrinos from the Spallation Neutrino Source~(SNS).
Later on, the $\nu$-Ar coherent scattering is also detected by the COHERENT experiment using a single phase liquid argon detector~\cite{Akimov:2020pdx}.
Since the neutrinos are produced in the pion-decay-at-rest process, the nuclei recoil energy concentrates in a range of several to tens of keV~(keV$_{\rm nr}$).

Given the high cross section and the unique properties of the CE$\nu$NS process, many studies have been carried out using the COHERENT data.
For example, the weak mixing angle is measured at the low momentum transfer~\cite{Cadeddu:2018izq}.
The average CsI neutron density distribution is obtained~\cite{Cadeddu:2018rlm}.
When endowed with mass and electromagnetic properties, the neutrino magnetic moment $\mu_{\nu}$~\cite{Kosmas:2017tsq} and the neutrino charge radius~\cite{Cadeddu:2018dux} are limited.
On the other hand, via the CE$\nu$NS process, astrophysical neutrinos create an unavoidable background in the experiments searching for the WIMP dark matter~\cite{Aprile:2016pmc,Zhang:2018xdp,DarkSide20k}, that is the 'neutrino floor'.
In addition, the search for low-mass WIMP using the ionization signal only has drawn more and more attention~\cite{Agnes:2018ves,Aprile:2019jmx,Akerib:2020aws}.
Understanding the response of LXe and LAr from sub-keV to several keV nuclear recoil energies is crucial to the search for light dark matter via this approach.

Given such abundant physics topics, many experiments are on projecting or proceeding~\cite{Papoulias:2019xaw}.
Larger detectors with lower detection threshold will be employed in the COHERENT experiment~\cite{jason_newby_2020_4134026}.
On the other hand, experiments using the strongest artificial neutrino source on the Earth, i.e., the nuclear reactors, are growing quickly, such as CONNIE~\cite{Aguilar-Arevalo:2019jlr}, MINER~\cite{Agnolet:2016zir}, CONUS~\cite{Bonet:2020awv}, and Red-100~\cite{Akimov:2019ogx}.
Compared to the neutrinos with energies of a few tens MeV at SNS, the 2~MeV average energy of reactor antineutrinos requires a significant reduction of the detection threshold.
Thus, the corresponding detectors used in the above experiments are the cryogenic Ge/Si detector, or the dual-phase Xe Time Projection Chamber~(TPC).

In this paper, we propose an experiment to measure the coherent scattering between reactor \nuebar's and argon nuclei using a dual-phase argon TPC at Taishan, referred to as Taishan Coherent Scattering Experiment here after.
The Taishan Coherent Scattering Experiment will use 200~kg argon as target and will be located side-by-side with the JUNO-TAO experiment~\cite{Abusleme:2020bzt}, which aims to precisely measure the reactor \nuebar~spectrum using a ton-scale liquid scintillator detector with a percent level energy resolution.
The antineutrino flux is approximately 6$\times10^{12}$~cm$^{-1}$s$^{-1}$ at this location, and the vertical overburden is about 5~meters of water equivalent~(m.w.e.)~\cite{Abusleme:2020bzt}.
The TPC size is limited primarily by the size of the elevator to enter the lab.
Thus, the maximum fiducial argon mass is 200~kg.

There are many advantages of using argon in the detection of CE$\nu$NS.
As a noble element detector, the target mass is easy to reach hundreds of kilograms.
This leads to much larger statistics compared to the kg-scale Ge/Si detectors.
Although the $\nu-$Xe CE$\nu$NS cross section is about 16 times larger than that of $\nu-$Ar, due to the mass of an argon nucleus is approximately 3.4 times smaller than xenon, the larger nuclear recoil energy leads to much easier detection in argon.
The large CE$\nu$NS sample not only allows to measure the weak mixing angle $\sin\theta_W$ using the reaction rate, but also opens a door to explore more physics using the spectral shape, for example, constraining the neutrino magnetic momentum, and measuring the ionization electron yield at keV$_{\rm nr}$ energy range.
The major problem of using argon is the intrinsic 1~Bq/kg $^{39}$Ar decays in the atmospheric argon~(AAr).
A possible solution is to use the depleted underground argon~(UAr)~\cite{Agnes:2015ftt}.
As a feasibility study, we will use a 0.7~mBq/kg $^{39}$Ar decay rate in the following sections.

The paper is structured as follows:
Sec.~\ref{layout} presents the conceptual design of the low threshold Argon TPC detector.
Sec.~\ref{background} focuses on the background simulation and the design of shielding.
Sec.~\ref{sensitivity} presents the study of expected sensitivities.
A short summary and prospects are discussed in Sec.~\ref{summary}.

\section{Conceptual design of Taishan Coherent Scattering Experiment}
\label{layout}

In the Standard Model, the differential cross section of CE$\nu$NS is expressed as:
\begin{equation}
\label{eq:sigma}
   \frac{d\sigma}{dT_{N}}\left ( E_{\nu}, T_{N}\right )\simeq \frac{G_{F}^{2}}{4\pi}MQ^2_W
   \times \left( 1-\frac{MT_{N}}{2E_{\nu}^{2}} \right )F^{2}_W(q^2),
\end{equation}
where $E_{\nu}$ is the energy of the neutrino, $G_F$ is the Fermi constant, $M$ is the nucleus mass, and $T_N$ is the nuclear recoil energy. The weak charge $Q_W$ is defined as:
\begin{equation}
\label{eq:qw}
Q_W = N-(1-4\times\sin\theta_W)\cdot Z,
\end{equation}
where $N$ and $Z$ are the numbers of neutrons and protons in the nucleus, and $\theta_W$ is the Weinberg mixing angle.
The weak form factor $F^{2}_W(q^2)$ is taken as 1 due to the low $q$ values in the scattering between reactor \nuebar's and nuclei~\cite{Payne:2019wvy}.

\begin{figure}[h]
  \begin{center}
  \includegraphics[width=0.7\textwidth]{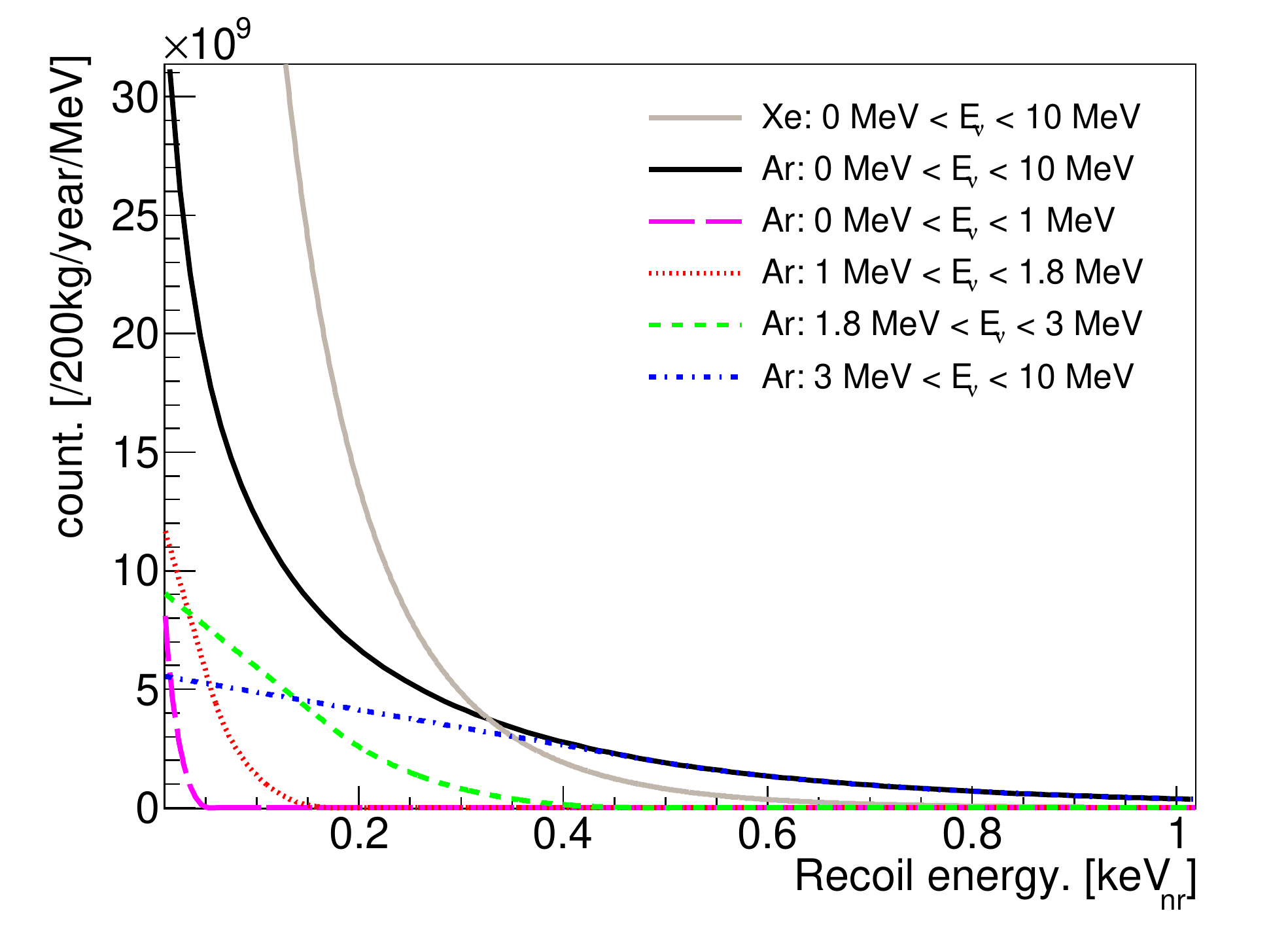}
  \caption{\label{fig:Spectrum} Recoil energy spectra of Ar and Xe nuclei in the reactor \nuebar~CE$\nu$NS reaction.}
  \end{center}
 \end{figure}

Using the reactor neutrino spectra in the 2-8 MeV range provided in Refs.~\cite{Mueller:2011nm,Huber:2011wv} and applying an exponential extrapolation for spectra below 2~MeV, the nuclear recoil energy spectra of Ar are calculated and shown in Fig.~\ref{fig:Spectrum}.
%
%
The \nuebar-Ar CE$\nu$NS reactions from neutrinos in sliced energy ranges~(0-1~MeV; 1-1.8~MeV; 1.8-3~MeV; 3-10~MeV) are also drawn for better illustration.
The detection of reactor \nuebar's with energies smaller than 1.8~MeV, i.e., below the threshold of the inverse-beta-decay reaction, is quite difficult.
Because the nuclear recoil energy is less than 0.15~keV$_{\rm nr}$ where the electron yield is about ${\rm 6.4~e^{-}/keV_{nr}}$ , corresponding to a single ionization electron.
In the energy region above 0.2~keV$_{\rm nr}$, the signals are dominated by \nuebar's with energies larger than 3~MeV.
This puts a stringent requirement on lowering the detection threshold to sub-keV$_{\rm nr}$.

The dual-phase TPC using argon or xenon suits such low energy detection.
An obvious advantage of argon is the smaller nucleus mass than xenon.
The signal rate on argon is larger than that on xenon with the same target mass when the recoil energy is larger than 0.17~keV$_{\rm nr}$.
To estimate a feasible detection threshold, we take the experience from the DarkSide-50 and the Red-100 experiments.
The DarkSide-50 dual-phase argon TPC has achieved the detection of single ionization electron and the analysis of energy threshold above 4 ionization electrons.
The latter value corresponds to approximately 0.5~keV$_{\rm nr}$~\cite{Agnes:2018ves}.
The Red-100 experiment has performed the first ground-level laboratory test of the dual-phase xenon TPC and proved the feasibility of a threshold of 4 ionization electrons~(about 0.6~keV$_{\rm nr}$ in xenon)~\cite{Akimov:2017hee, Akimov:2019ogx,Khaitan:2018wnf}.
Thus, we will use 4 ionization electrons~(about 0.5~keV$_{\rm nr}$ in argon) as the analysis threshold in the following studies.

The proposed Taishan Coherent Scattering Experiment will be located side-by-side with the JUNO-TAO experiment in the Taishan Nuclear Power Plant.
There are two reactor cores in operation now.
Each of them has a thermal power of 4.6~GW.
The experiment will be in a basement of an elevation of –9.6 meters, in which the measured cosmic-ray muon flux is one third of that at the ground level.
The argon TPC is approximately 35~m to one reactor core and is 253~m to another one.
The entrance to the laboratory relies on an elevator with a width of 1.39~m and a height of 1.99~m.
Thus, the size of TPC together with the outer shielding and insulation materials is limited to 1.4~m and the total fiducial mass is about 200~kg.
In addition, the height of the detector, including the shield, should be less than 3.85~m due to the limitation of experimental hall space.

\begin{figure}[h]
  \begin{center}
  \includegraphics[width=0.7\textwidth]{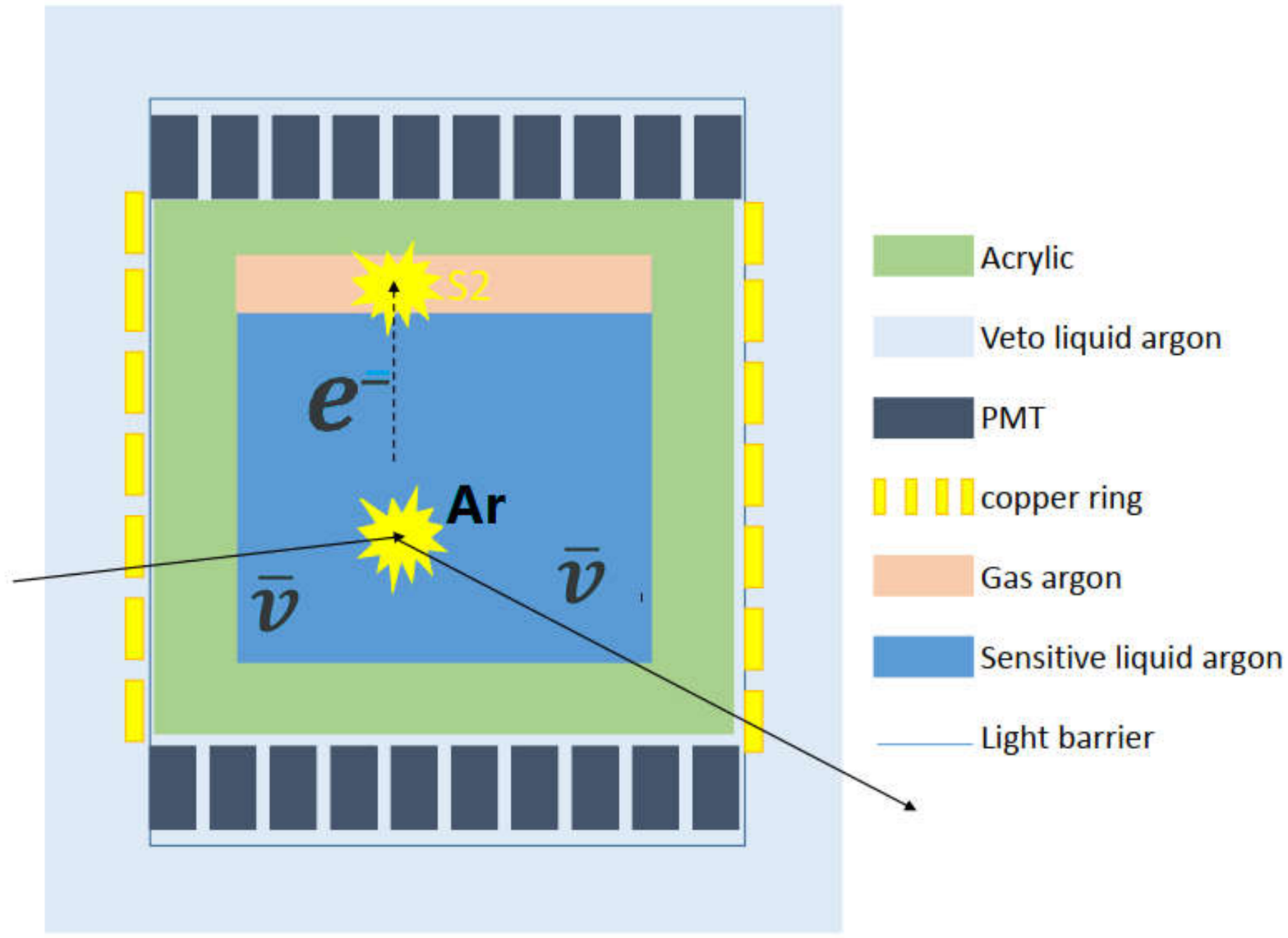}
  \caption{\label{fig:TPC} Conceptual design of the dual-phase argon TPC at Taishan. The veto liquid argon detector is of great importance to tag the multiple scattering $\gamma$'s and neutrons.}
  \end{center}
\end{figure}

Figure~\ref{fig:TPC} shows the conceptual design of the dual-phase argon TPC detector.
The most inner region is the sensitive argon target~(SensAr).
A cylindrical acrylic vessel~(57~cm height, 57~cm diameter) is used to hold ${\rm \sim 200~kg}$ liquid argon.
A drift electric field is applied in the vertical direction by the high votage on the copper rings.
A gas argon layer with a height of 1.5~cm is above the liquid argon.
A luminescent electric field is applied in the gas region for the single ionization electron detection.
The energy deposition in liquid argon produces photons and free electrons.
The photons are promptly detected by photosensors and are commonly named as S1.
The free electrons will drift to the gas layer in the vertical electric field.
Then, one electron could generate an electroluminescence signal named as S2.
The S2 photoelectron (p.e.) yield can reach 23$\pm$1 ${\rm p.e./e^{-}}$~\cite{Agnes:2018ves}.
The time between S1 and S2 is the drift time corresponds to the time it takes for electrons to move from the action point to the gas phase. 
The electron drift speed is related to the strength of the drift electric field~\cite{Carugno:1990kd}.
%
In the reactor \nuebar-Ar CE$\nu$NS reaction, it is great difficulty to detect S1.
For the argon detector, the better light yield is about ${\rm 10~PE/keV_{ee}}$ for S1.  So the S1 is few and even zero PE in the ${\rm sub-keV_{nr}}$ considering the quenching factor. 
Thus, the experiment would rely on S2 only.
A set of photosensor array is equipped on the top and the bottom of the TPC.
The photon sensitive coverage is ${\rm 30\%}$ similar to that of DarkSide-50.
To simulate the background, 44*2 Hamamatsu R-11065 PMTs are used as the photon sensor array.

\section{Veto and shielding design}
\label{background}

\begin{figure}[h]
  \begin{center}
  \includegraphics[width=0.7\textwidth]{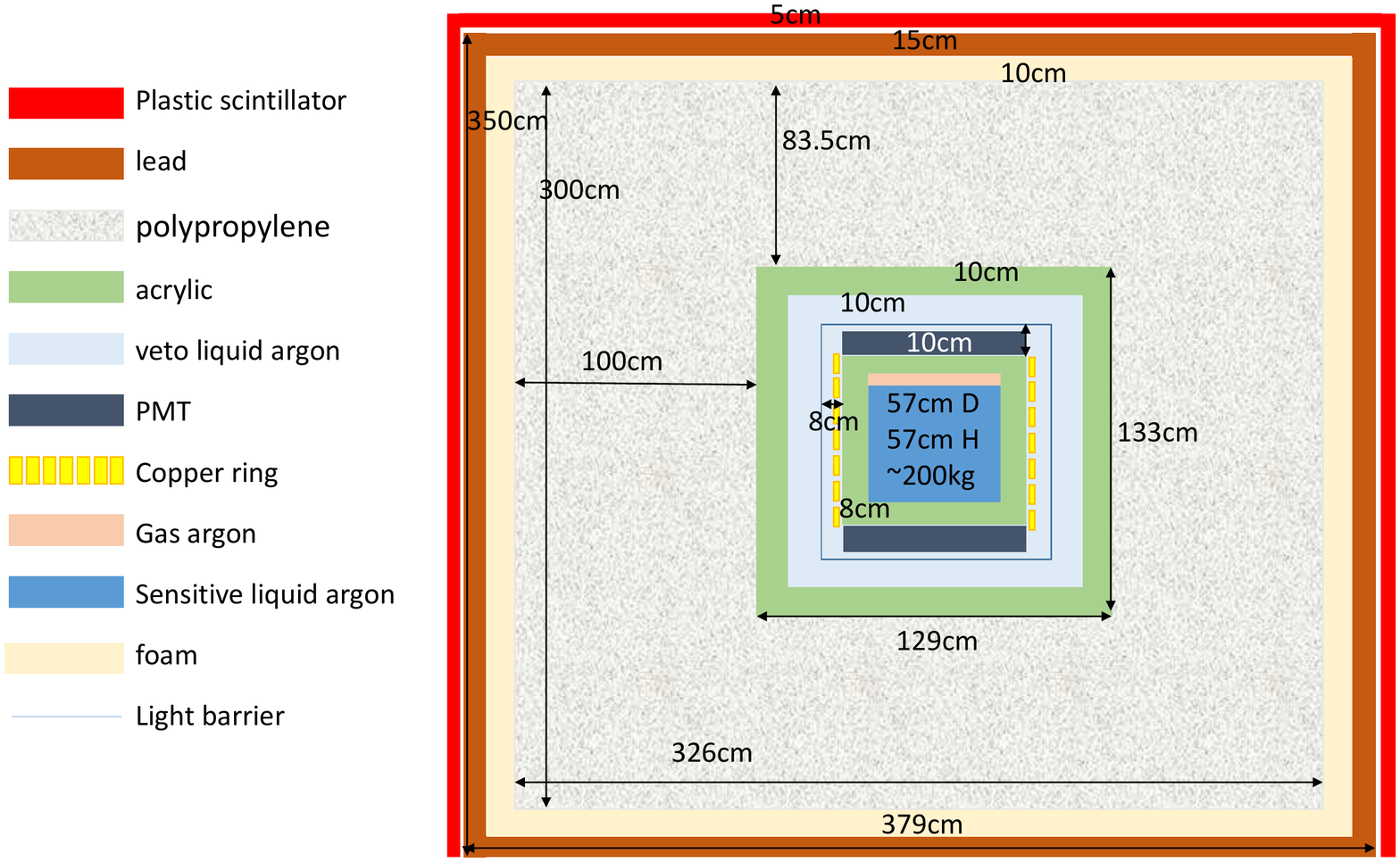}
  \caption{\label{fig:Neardetector} Conceptional design of veto and shielding of the Taishan Coherent Scattering Experiment.}
  \end{center}
\end{figure}

Although the number of CE$\nu$NS signals could reach 1,000 per day in the 200~kg argon target above a detection threshold of 4 ionization electrons, the signals could be easily washed out by the backgrounds from cosmic ray muons~(about 60~Hz/m$^2$) and ambient radioactivity decays.
To design the veto and shielding, lots of simulation are carried out using Geant4~\cite{Geant4}.
The ambient radioactivity and cosmic ray muons are simulated.
The concentrations of $^{238}$U, $^{232}$Th, and $^{40}$K in the materials are listed in Table~\ref{tab1:impurities}.
The muon generators are taken from the JUNO-TAO simulation~\cite{Abusleme:2020bzt}.
The veto and shielding design is shown in Fig.~\ref{fig:Neardetector}.

\begin{table*}[h]
\centering
\begin{tabular}{cccccc}
\hline
material     & weight & Data source                        & ${\rm ^{238}U~mBq/kg}$       & ${\rm ^{232}Th~mBq/kg}$       & ${\rm ^{40}K~mBq/kg}$              \\
\hline
Lead         & 101~t   & radiopurity.org     & 60                        & \_                         & 14          \\
PP           & 32~t    & radiopurity.org       & 9.84                      &0.569                       & \textless{3.1} \\
Acrylic      & (814+190)~kg  & JUNO-TAO      & \textless{0.0123}         & ${\rm \textless{4.07\times 10^{-3}}}$   & ${\rm \textless{3.1\times 10^{-5}}}$   \\
Copper ring  & 40~kg   & DarkSide-50   & \textless{0.06}           & \textless{0.02}            &0.12         \\
3" PMT R11065 & ${\rm 44\times2}$ & Darkside-50             & 5.2~mBq/PMT               & 13.4~mBq/PMT               & 37.1~mBq/PMT           \\
concrete     & 400~t   & JUNO-TAO                          & ${\rm 5.8\times 10^{4}}$  &${\rm 7.9\times 10^{4}}$    &${\rm 7.8\times 10^{5}}$     \\
\hline
\end{tabular}
\caption{ \label{tab1:impurities} Concentrations of radioactive impurities of materials.}
\end{table*}

The outermost layer is plastic scintillator with 5~cm thickness to tag cosmic ray muons.
%
%
The inner brown layer stands for 15~cm lead to shield the ambient radioactivity from hall.
Then, a polypropylene layer acts as thermal insulation and also can shield neutrons generated by muons.
%
%
With 90~cm polypropylene, the acrylic vessel can keep -186~$^{\rm o}$C stably.

%
There are two liquid argon detectors, the outer liquid argon detector is served as a veto detector, instrumented with photosensors and named as the VetoAr, and the inner detector is the aforementioned dual-phase liquid argon TPC, used to detect \nuebar's and named as the SensAr. 
The VetoAr not only passively shields the SensAr from radioactivity from outer materials or hall concrete, but also can tag the multiple scattering $\gamma$s or neutrons.
Once the VetoAr observes an energy deposit, a 600~$\mu$s veto window could be applied to the SensAr according to the maximum drift time of electrons from the TPC bottom to the top. 
Since most of backgrounds are from gammas, a quenching factor of 0.25~\cite{Gastler:2010sc} is used to convert the nuclear recoil energy and the electron energy, that is, 1~keV$_{\rm ee}$ = 4~keV$_{\rm nr}$.
In the following sections, the backgrounds are provided in the energy range of 0 to 1~keV$_{\rm ee}$, corresponding to 0 to 4~keV$_{\rm nr}$.

\subsection{Natural radioactivity}

\subsubsection{Backgrounds from the hall}

\begin{figure}[h]
  \begin{center}
  \includegraphics[width=0.7\textwidth]{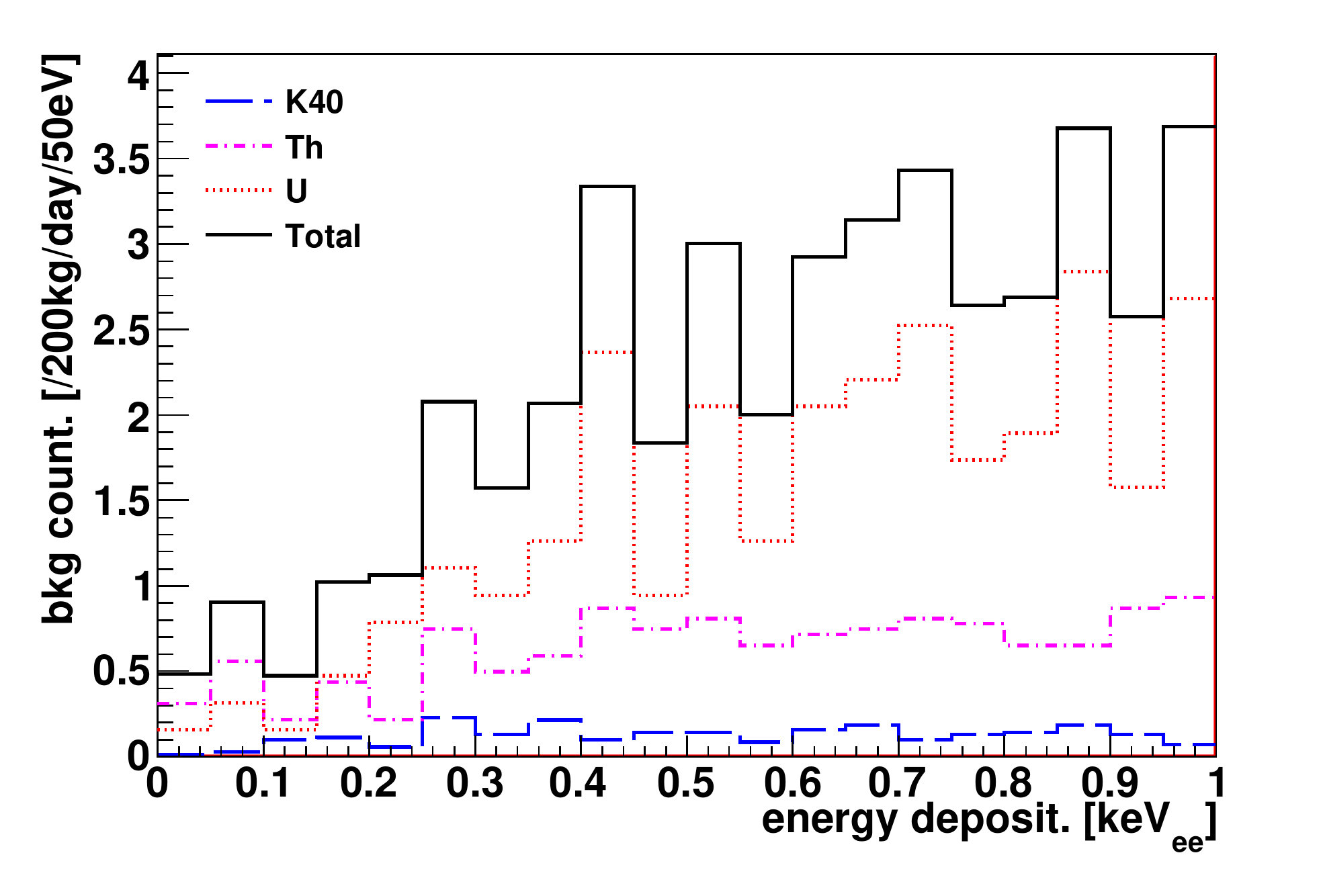}
  \caption{\label{fig:HallSPectrum} The deposit energy spectrum in the SensAr from natural radioactivity decays in the hall concrete after shielded by 15~cm lead and other detector materials.}
  \end{center}
\end{figure}

In the simulation of the natural radioactive decays in the concrete of the hall, the simulation generators including the ${\rm \gamma/\alpha/e^{-}}$ from radioactive decay are placed in the shell extending 30~cm into the inner wall.
According to simulation results, the gamma rays beyond the 30~cm are almost impossible to transport into the sensitive argon.
Figure~\ref{fig:HallSPectrum} shows the energy spectrum in the SensAr after shielded by 15~cm lead and other inner materials.
The background level is about ${\rm 0.22~/day/kg/keV_{ee}}$ in the 0 to 1~keV$_{\rm ee}$ energy region.
Considering the 101~t weight of the 15~cm lead, different thicknesses of lead are also simulated, listed in Table~\ref{tab:HallLead}.
It can be found that the 15~cm lead is difficult to reduce.

\begin{table}[h]
\begin{center}
\begin{tabular}{cccc}
\hline
Lead Thickness.   & 15~cm & 13~cm & 10~cm \\
\hline
Background & 0.22 & 0.91 & 5.37 \\
\hline
\end{tabular}
\caption{\label{tab:HallLead} Background level~(/day/kg/keV$_{\rm ee}$) in the SensAr from the hall concrete after shielded by different thickness of lead.}
\end{center}
\end{table}

In addition to the gammas, fast neutrons could be produced via the $(\alpha,n)$ reaction and in the $^{238}$U spontaneous fission, named as radiogenic neutrons.
The neutron flux from Hall is estimated as ${\rm 2.0 \times 10^{-6}~/cm^{2}/s}$~\cite{Bungau:2005xp}, and the average neutron kinetic energy is about 1.7~MeV.
The attenuation length of neutron is about 6.8~cm in the material that contains ${\rm 10\%}$ hydrogen.
After about ${\rm 90~cm}$ polypropylene, the neutron flux could be reduced by ${\rm 10^{6}}$.
Thus, the radiogenic neutron contributes only a few background per day that can be ignored.
%

  \subsubsection{Backgrounds from the detector material}

\begin{figure}[h]
  \begin{center}
  \includegraphics[width=0.7\textwidth]{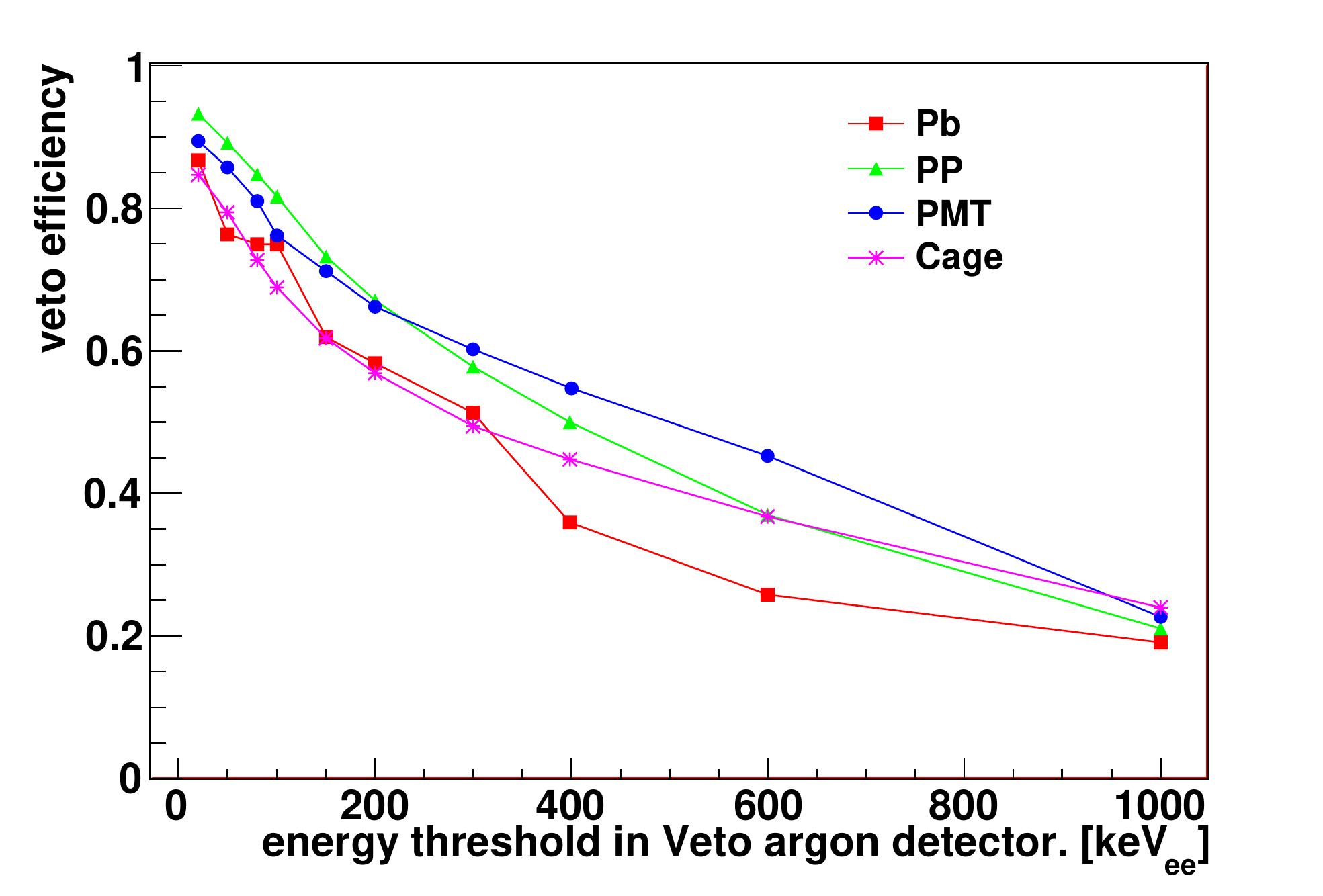}
  \caption{\label{fig:VetoEfficiency} Veto efficiency with respect to the energy threshold in the veto argon detector}
  \end{center}
\end{figure}

The background contributions in the SensAr from major detector components are listed in Table~\ref{tab1:bkgnatural}.
If the energy deposit in the VetoAr is larger than 0.1~MeV, this event triggers the veto argon detector, and a subsequent 600~$\mu$s veto window is applied in the sensitive detector.
Here list the statistics of radioactivity counts ${\rm ( E_{\rm dep}~\textgreater{}~0.1~keV)}$ and also list the count of the energy interval ${\rm (0.1~keV~\textless{}~E_{\rm dep}~\textless{}~1~keV)}$. The radioactivity from material could be vetoed by veto argon detector.
The threshold in the veto detector is also varied, and the corresponding veto efficiency is shown in Fig.~\ref{fig:VetoEfficiency}.
The different color mean different material.
For higher veto efficiency and lower trigger, an energy threshold of 0.1~MeV is recommended. 
%
The "${\rm E_{dep}}$" is the energy deposit in the sensitive argon.
The total trigger rate will be about 2~Hz in the sensitive detector.
With the help of the VetoAr, the background in the energy range of interest is only 18 per day in SensAr, much smaller than the 1,000 signals.

  \begin{table*}[h]
  \footnotesize
    \begin{tabular}{|c|c|c|c|c|c|c|c|c|}
    \hline
    \multicolumn{2}{|c|}{Material}          & Lead   & PP & OutAcrylic & InnerAcrylic & Cage   & PMTs   & Total      \\ \hline
    E$_{\rm dep}$~\textgreater{}~0.1~keV     &       & 34753  & 60060         & 40        & 251           & 28580  & 50596  & 174280 \\ \cline{1-1} \cline{3-9}
    0.1~keV~\textless{}~E$_{\rm dep}$~\textless{}~1~keV & \multirow{-2}{*}{No veto}
                                            & 11.2 & 25.9         & 0           & 0.14          & 10.4  & 26.5  & 74.2    \\ \hline
    E$_{\rm dep}$~\textgreater{}~0.1~keV     &       & 14576  & 19658         & 5.5         & 172         & 13357  & 14990  & 62758  \\ \cline{1-1} \cline{3-9}
    0.1~keV~\textless{}~E$_{\rm dep}$~\textless{}~1~keV & \multirow{-2}{*}{After veto}
                                            & 4.1   & 4.7          & 0           & 0.1        & 3.3  & 5.6   & 17.8    \\ \hline
    \multicolumn{2}{|c|}{Dead time}
                                            & ${\rm 1.1\%}$ & ${\rm 1.4\%}$        & ${\rm 0.0\%}$      & ${\rm 0.25\%}$        & ${\rm 0.1\%}$ & ${\rm 0.2\%}$ & ${\rm 3.0\%}$   \\ \hline
    \end{tabular}
    \caption{\label{tab1:bkgnatural} Background rates in the SensAr per 200~kg per day. The veto means a larger than 0.1~MeV energy deposit is found in the VetoAr, then, a subsequent 600~$\mu$s veto is applied in the SensAr.}

  \end{table*}

Another crucial background comes from the intrinsic $^{39}$Ar and $^{85}$Kr decays in the argon.
It is well known that the $^{39}$Ar concentrations in the argon from atmosphere~(AAr) or underground~(UAr) are quite different as shown in Table~\ref{tab1:ArKr}~\cite{Agnes:2015ftt,Agnes:2014bvk}.
If the AAr is used in the SensAr, the number of $^{39}$Ar decays in the energy of interest could reach 40 times of the signal.
Thus, UAr is a must for the detection of CE$\nu$NS using reactor \nuebar's.
In this case, the background rate is 25 per day per 200~kg in the energy range of interest.
If the AAr is used in VetoAr, the dead time could reach more than 50\%.

\begin{table}[h]
\footnotesize
\begin{center}
\begin{tabular}{c|c}
\hline
 $^{39}$Ar in underground Ar & 0.73$\pm$0.11~mBq/kg \\ \hline
 $^{39}$Ar in atmospheric Ar & (1.01$\pm$0.08)$\times 10^3$~mBq/kg \\ \hline
 $^{85}$kr in Ar & 2.05$\pm$0.13~mBq/kg \\ \hline
\end{tabular}
\caption{\label{tab1:ArKr} Intrinsic radioactivity in liquid argon. The data are taken from Refs.~\cite{Agnes:2015ftt,Agnes:2014bvk}.}
\end{center}
\end{table}

\begin{figure}[h]
  \begin{center}
  \includegraphics[width=0.45\textwidth]{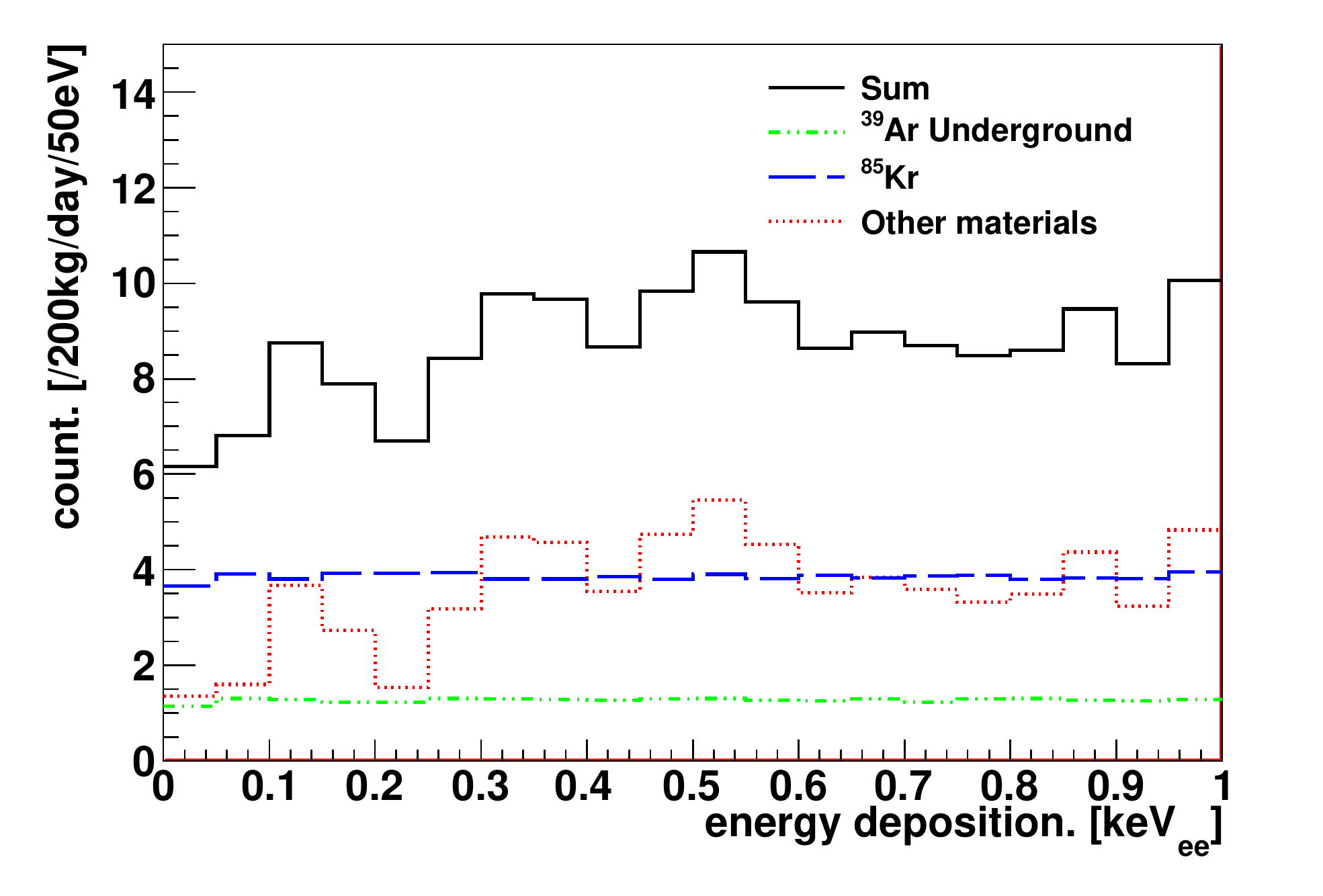}
  \includegraphics[width=0.45\textwidth]{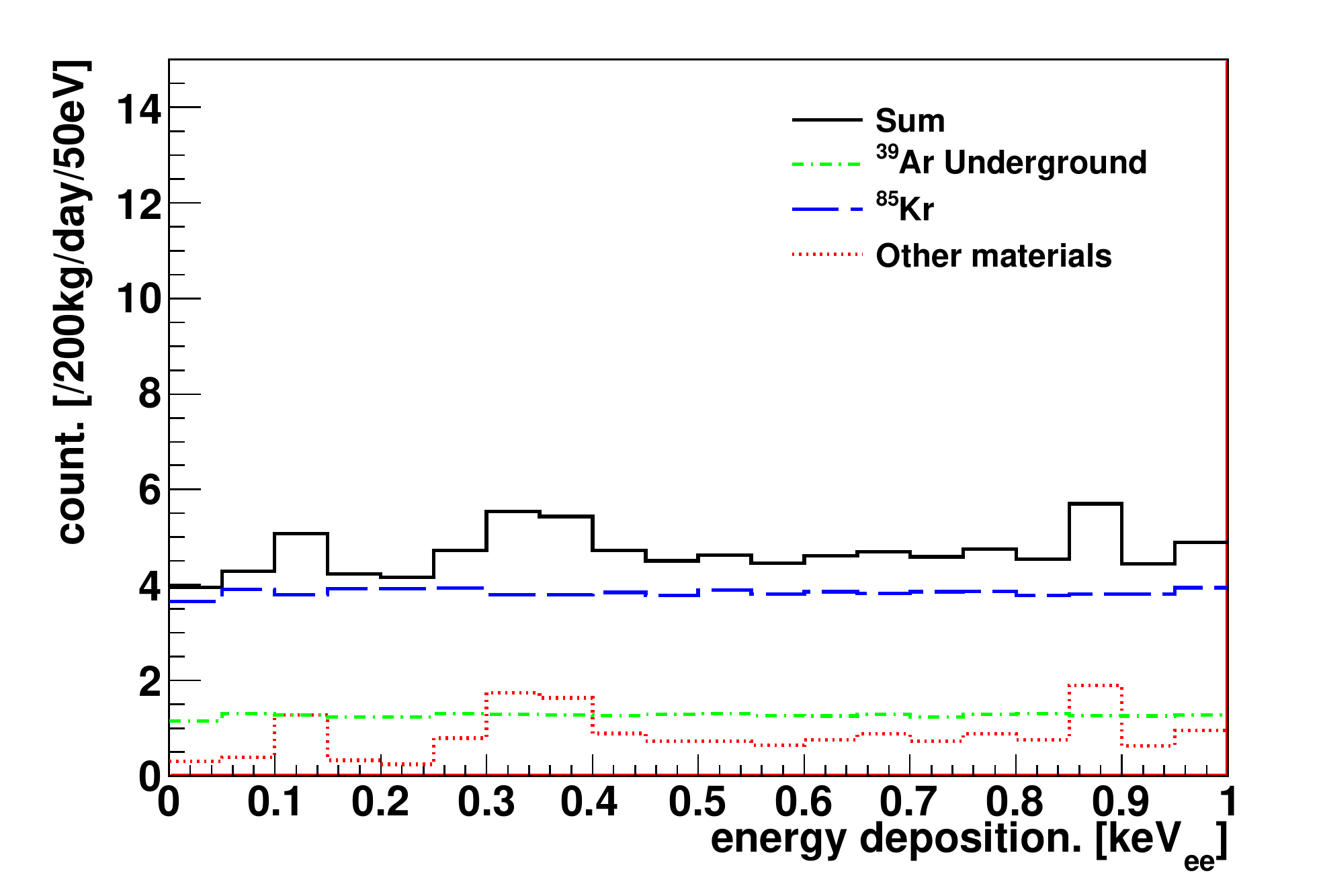}
  \caption{\label{fig:AfterNatural} Backgrounds from materials before~(left) and after~(right) implementing the veto of VetoAr.}
  \end{center}
  \end{figure}

As a short summary, the energy spectra of backgrounds from materials are shown in Fig.~\ref{fig:AfterNatural}.
%
%
The residual backgrounds are dominated by the intrinsic $^{39}$Ar and $^{85}$Kr decays.
A lower $^{85}$Kr concentrations could be achieved using the distillation at low temperature.

 \subsection{Cosmic ray muons}

\begin{figure}[h]
  \begin{center}
  \includegraphics[width=0.7\textwidth]{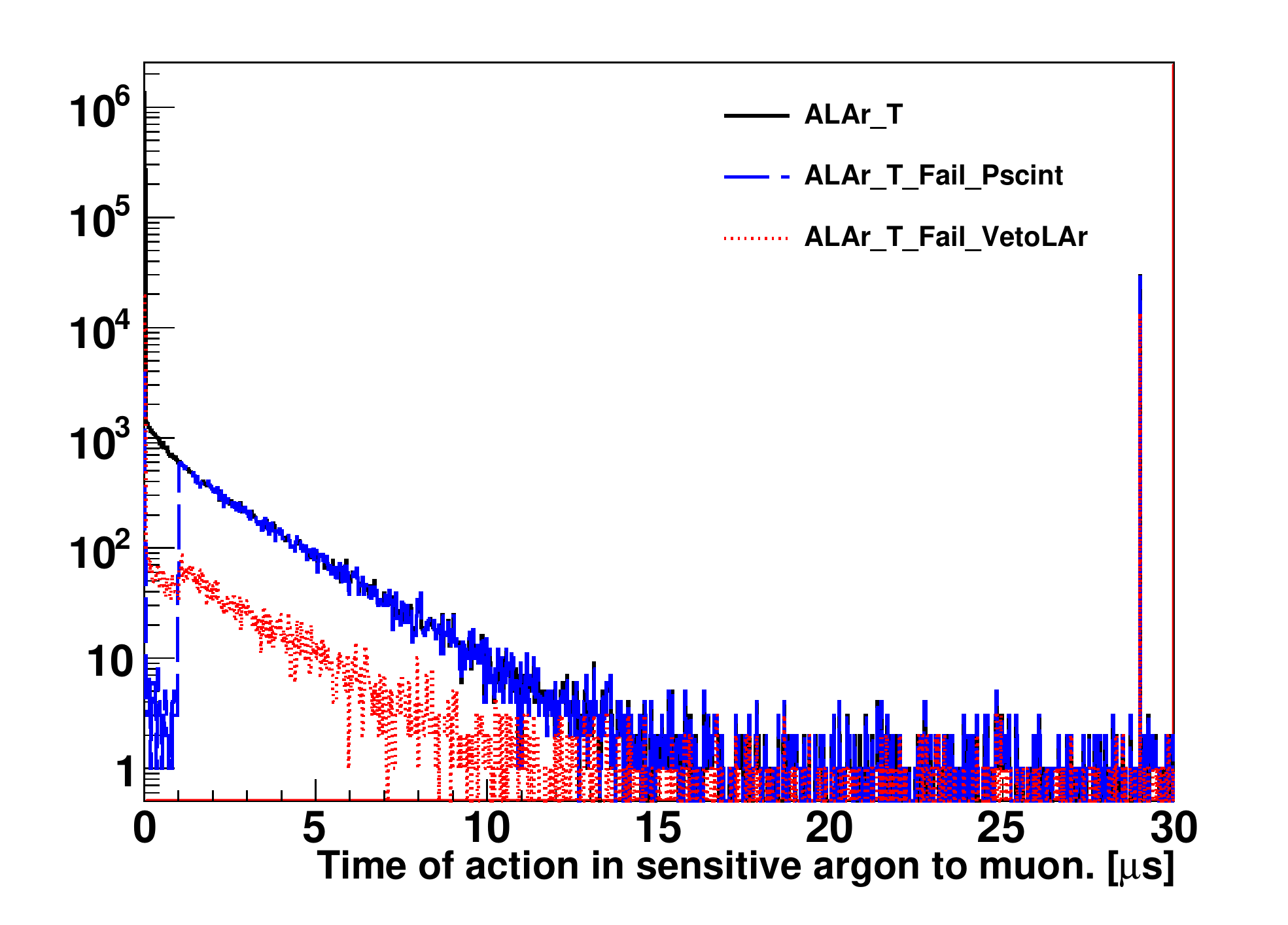}
  \caption{\label{fig:TimeArgon} Time distribution to last muon of energy deposit in SensAr. Black line shows the all events in SensAr; blue line shows the residual events after the veto of plastic scintillator; red line shows the residual events after the veto of VetoAr.}
  \end{center}
 \end{figure}

Although most of the hadronic components of cosmic rays have been absorbed, the muons with a rate of $\sim$60~Hz/m$^2$ in the experimental hall would contribute significant backgrounds if they are not well tagged.
In the conceptual design, two detectors are employed to tag the muons.
The first one is a layer of plastic scintillator with 5~cm thickness as the outmost tag.
The VetoAr detector is used to tag muons close to the sensitive argon TPC as Fig.~\ref{fig:Neardetector}.
Table~\ref{tab1:rate} summarizes the rates and veto efficiency of muons and secondary particles in different detectors.
Figure~\ref{fig:TimeArgon} shows the time interval distribution between the energy deposit in SensAr and the corresponding muon.
Black line shows the all events in SensAr; blue line shows the residual events after the veto of plastic scintillator; red line shows the residual events after the vetof of VetoAr.
For illustration, the time interval longer than ${\rm 30~\mu s}$ is filled in the bin of ${\rm 29~\mu s}$.
There are two typical cases to discuss.

\begin{table}
\footnotesize
\begin{center}
\begin{tabular}{cccc}
\footnotesize
\\ \hline
Detector & Plastic   & Veto  &Sensitive  \\
          & scintillator  &argon  &argon \\\hline
Rate.~Hz  & 3421  & 325     & 106  \\ \hline
veto efficiency.~\%  & 95.7\%  & 97.8\%     & -  \\ \hline
\end{tabular}
\caption{ \label{tab1:rate} Rates and veto efficiency of cosmic ray muons and their secondary particles in different detectors.}
\end{center}
\end{table}

{\textbf I.}  Muons passing through or stopped in the SensAr or VetoAr.
Once a large energy deposit is found in the two detectors, a subsequent 600~$\mu$s veto window is applied in the SensAr.
Neutrons or radioactive nuclei could be produced in the interaction of cosmic ray muon and detector materials, such as radioisotopes of S/Cl/Ar/K.
The energetic particles generated in the neutron capture and the nuclei decays form the long tail in the time interval distribution.
The 600~$\mu$s veto window removes most of neutrons and muon decay signals.
The events with longer life are difficult to be vetoed.
The VetoAr could reduce significantly such background because a $\gamma$ from the decay of cosmogenic isotopes could deposit energy both in the VetoAr and SenAr detectors, shown as the red line in Fig.~\ref{fig:TimeArgon}.

{\textbf II.} For muons only passing through the plastic scintillator detector, the plastic scintillator could tag them with a very high efficiency.
Due to the 3400~Hz trigger rate of the plastic scintillator detector, the subsequent correlation time window with the argon detectors could set only to 1 to 10~$\mu$s.
It could help to classify the events in the argon detectors, as cosmic ray muons contribute the largest trigger rate in the argon detectors, as listed in Table~\ref{tab1:rateSum}.

\subsection{Summary of background rates and the detector live time}

\begin{table*}
\begin{center}
\footnotesize
\begin{tabular}{c|c|c|c|c}
\hline
    & Concrete of Hall & Material       & Muon    & Total              \\ \hline
Rate in SensAr~(\textgreater{}~0.1~keV$_{\rm ee}$) & 1.1~Hz  & 2.6~Hz    & 106~Hz   & 109.7~Hz          \\ \hline
Rate in VetoAr~(\textgreater{}~0.1~MeV$_{\rm ee}$)& --    & 53~Hz+(711~Hz)   & 325~Hz   & 378~Hz+(711~Hz)    \\ \hline
Dead time   & --  & ${\rm 3.2\%~+~(42\%)}$  & ${\rm 19.5\%}$ & ${\rm 22.5\%~+~(42.0\%)}$ \\ \hline
\begin{tabular}[c]{@{}c@{}}Bkg after veto in 0.1-1~keV$_{\rm ee}$  \\ /keV/kg/day\end{tabular}  & 0.22  & \begin{tabular}[c]{@{}c@{}}0.09~+~0.12~+~0.38  \\ Materials+$^{39}$Ar+$^{85}$Kr\end{tabular} & 0.61   & 1.42            \\ \hline
\end{tabular}
\caption{\label{tab1:rateSum} Summary of background rates, and the veto dead time. The numbers in the brankets stands for the case that the air argon is used in the veto argon detector.}
\end{center}
\end{table*}

Table~\ref{tab1:rateSum} lists the summary of background rates and dead time.
The veto time window is 600~$\mu$s in SensAr once the VetoAr observes an energy deposit larger than 0.1~MeV.
This design significantly reduces the background in the SensAr.
A 10$\mu$s correlation window will be used between the argon detectors and the plastic scintillator detector.
A crucial requirement is that the argon of SensAr must be low radioactivity underground one.
Using UAr in the VetoAr is not necessary, but it can save the 42\% dead time as the value in the brankets.
Moreover, the low trigger rate in VetoAr by using underground argon can help set up an online anti-coincidence to reduce the amount of data.
After shielding and veto, the background level is about 1.424~event/kg/keV/day in 0.1-1~keV$_{\rm ee}$.

\section{ Expected sensitivity of the Taishan Coherent Scattering Experiment}
\label{sensitivity}

With the $\sim$1,000 CE$\nu$NS signals per day of the Taishan Coherent Scattering Experiment above the 0.5~keV$_{nr}$ threshold, many physics topics can be studied, for example, the weak mixing angle at low momentum transfer and the neutrino magnetic moment, etc.
Moreover, the reactor \nuebar's, with 2\%$\sim$3\% spectrum shape precision measured in Ref.~\cite{An:2016srz}, provide a powerful tool to study the ionization electron yield at keV$_{\rm nr}$ energy range.
This will benefit the search of low-mass WIMP dark matter, in particular for the analyses using the ionization signal only~\cite{Aprile:2019xxb,Agnes:2018ves}.
A synergy between the neutrino detector and the dark matter detector is foreseen.
Some discussions are presented in this section.

\begin{figure}[h]
  \begin{center}
  \includegraphics[width=0.7\textwidth]{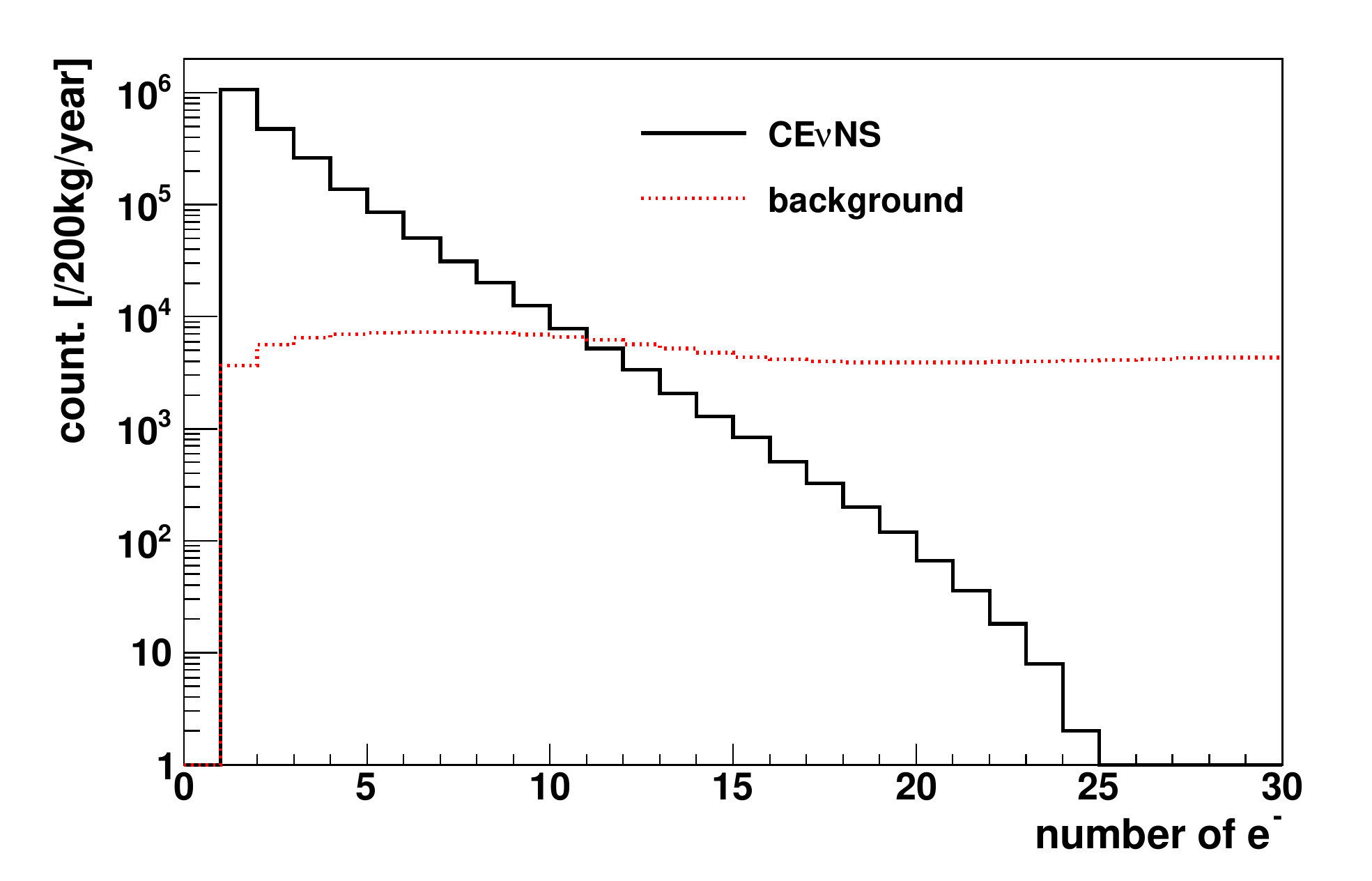}
  \caption{\label{fig:specee} Spectra of CE$\nu$NS signals and simulated background, taken a simplified detector response into consideration.}
  \end{center}
  \end{figure}

To perform these analyses, a simplified detector response is added to the signal and background spectra.
The first step is to convert the nuclear recoil energy and the electron energy to the number of ionization electrons~(N$_{e^-}$).
The ionization electron yield~(Y$_{e^-}$) refers to the results in Ref.~\cite{Agnes:2018ves}.
As mentioned above, a quenching factor of 0.25 is used between keV$_{\rm ee}$ and keV$_{\rm nr}$.
The N$_{e^-}$ distribution is assumed to follow the Poisson distribution.
The electron extraction efficiency is set to 100\%.
Figure~\ref{fig:specee} shows the CE$\nu$NS signal and background spectra in the unit of ionization electrons.
To get a good signal to background ratio, the energy region of interest is selected to 4 to 11~$e^-$.
The 4~$e^-$ is chosen to get rid of the backgrounds released by impurities after a large energy deposit.
In the 4 to 11~$e^-$ energy range, the number of signals and background events is 955 and 153, respectively, per day per 200~kg argon.

In the future, the background can be measured $in$-$situ$ using the reactor on-off information.
According to the reactor running status, there would be an one-month reactor off period per year.
The background can be measured to a statistic of N$_{\rm bkg}~=~153/{\rm day} \times 30~{\rm days} = 4590$.
This corresponds to a statistical uncertainty of $\sigma_{bkg}~=~1/\sqrt{N_{bkg}/N_{bin}}~=~4.2\%$ per ionization electron bin.
This number will be used in the following analyses.

The systematic uncertainties consist of the trigger efficiency, the target mass, and the reactor \nuebar~rate and spectrum.
According to Ref.~\cite{Agnes:2018ves}, the trigger efficiency has reached to 100\% at 4 ionization electrons.
Since all the sensitive argon is hold in the acrylic vessel, no fiducial volume cut is used, the target mass uncertainty only arises from the liquid level instability.
The \nuebar~prediction will benefit from the reactor neutrino experiments using the IBD channel~\cite{An:2016srz,Abusleme:2020bzt,Andriamirado:2020erz,AlmazanMolina:2020jlh}.
Thus, 1\%~(aggressive) and 3\%~(conservative, nominal) bin-to-bin correlated uncertainties will be tested.

  \begin{figure}[h]
  \begin{center}
  \includegraphics[width=0.7\textwidth]{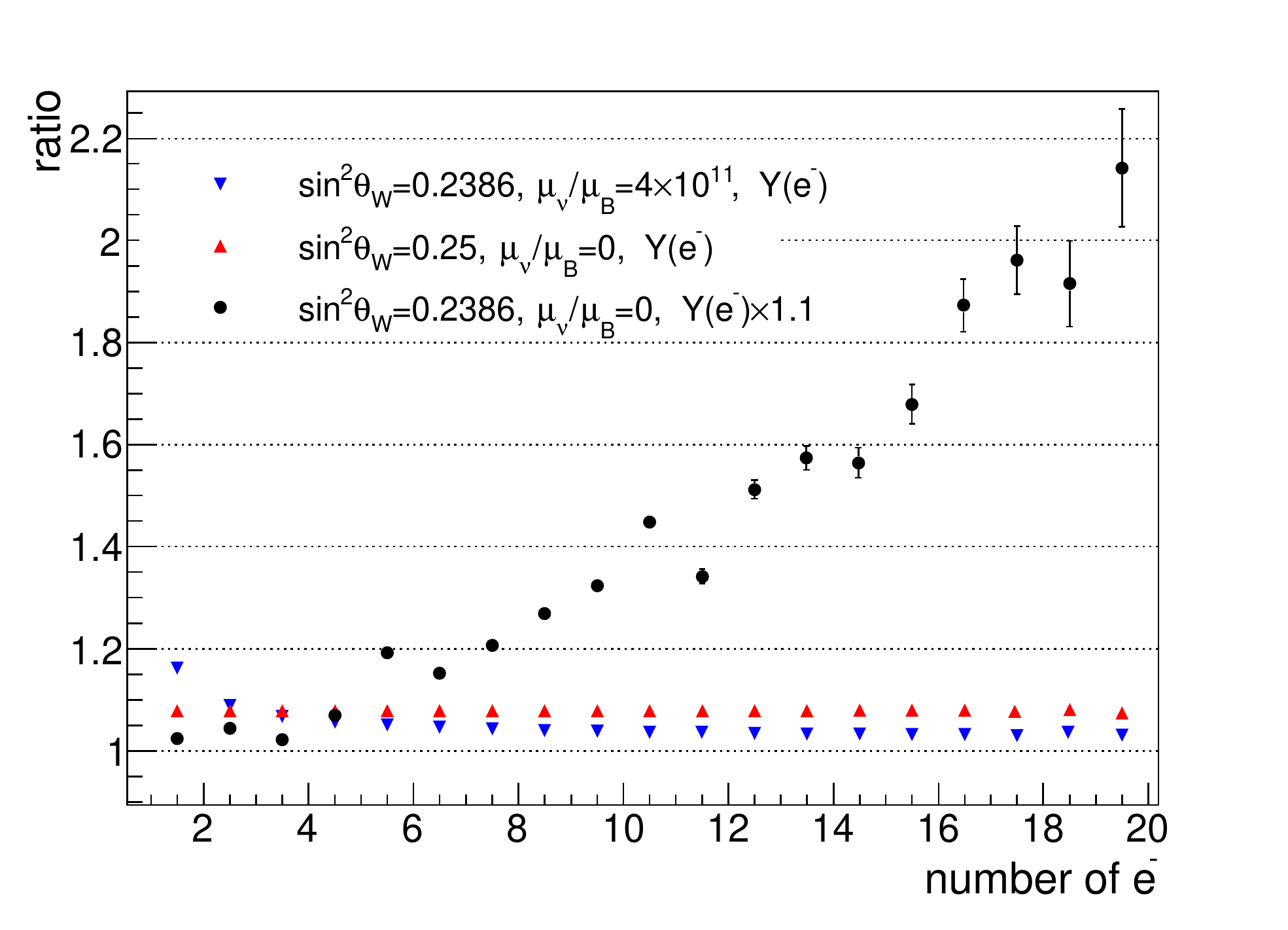}
  \caption{\label{fig:DiffSinMuYe} Ratios of spectra with different parameter values to norminal spectrum. Details are found in the text.}
  \end{center}
\end{figure}

The measured CE$\nu$NS spectrum in terms of the number of ionization electrons is sensitive to several parameters.
Here, we take the weak mixing angle $\sin^2\theta_W$, the neutrino magnetic moment $\nu_\mu$, and the ionization electron yield Y$_{e_-}$ for illustration.
The nominal value of $\sin^2\theta_W$ is set to 0.2386 according to the Particle Data Group~\cite{Zyla:2020zbs}, $\nu_\mu$ is set to 0, and Y$_{e^-}$ to 7 per keV$_{\rm nr}$.
Then, three spectra are generated by changing one parameter and keeping the other two unchanged.
Their ratios to the nominal spectrum are shown in Fig.~\ref{fig:DiffSinMuYe}.
The weak mixing angle $\sin^2\theta_W$ changes the number of events in all the bins by the same ratio.
An obvious upward trend is found when Y$_{e^-}$ is enlarged by 10\%.
It means the Y$_{e^{-}}$ can be extracted by fitting the measured spectrum shape with the high statistics data.
This provides a unique probe to calibrate the ionization electron yield in sub-keV$_{\rm nr}$ energy range, in which there is no suitable neutron source up to now.
Finally, if a non-zero $\nu_\mu$ value is introduced, a downward trend is found, since the CE$\nu$NS cross section will be plus the term:
\begin{equation}
    \label{eq:sigmamu}
   {\rm \frac{d\sigma _{\mu_{\nu}}}{dT_{N}}\left ( E_{\upsilon }, T_{N}\right )=\frac{\pi\alpha ^{2}}{m_{e}^{2}}\left | \frac{\mu_{\nu}}{\mu_{B}} \right |^{2}\frac{Z^{2}}{T_{N}}\left ( 1-\frac{T}{E_{\nu}}+\frac{T^{2}}{4E^2_{\nu}} \right )}.
\end{equation}

To obtain a expected sensitivity for the parameters, a $\chi^{2}$ function is defined as:
\begin{equation}
 \label{eq:chi}
{\rm \chi ^{2} = \underset{\alpha ,\beta }{\min}\sum_{i=4e^{-}}^{i=11e^{-}}\left(\frac{\left ( N_{data}^{i}-N_{Fit}^{i}\left ( 1+\alpha  \right )-Bkg^{i} \right )^{2}}{N_{data}^{i} + \beta^i\times Bkg^{i}} \right )}
{\rm +\left ( \frac{\alpha }{\sigma _{\alpha }} \right )^{2}},
 \end{equation}
where ${\rm N_{data}^{i}}$ is the measured candidates in the $i$th bin, and ${\rm N_{data}^{i}}$ is the predicted number of CE$\nu$NS signals.
The exposure is set to 200~kg$\cdot$year.
The ${\rm Bkg^{i}}$ is the number of background in the $i$th bin, and can be estimated using the reactor off data as mentioned above.
Thus, the 4.6\% background uncertainty is treated as bin-to-bin uncorrelated as $\beta^i$ in the denominator.
The nuisance parameter $\alpha$ stands for the systematic uncertainty, and 3\% is used conservatively as the nominal value.
The results with 1\% systematic uncertainty are also shown in the plots.

\begin{figure}[h]
  \begin{center}
  \includegraphics[width=0.45\textwidth]{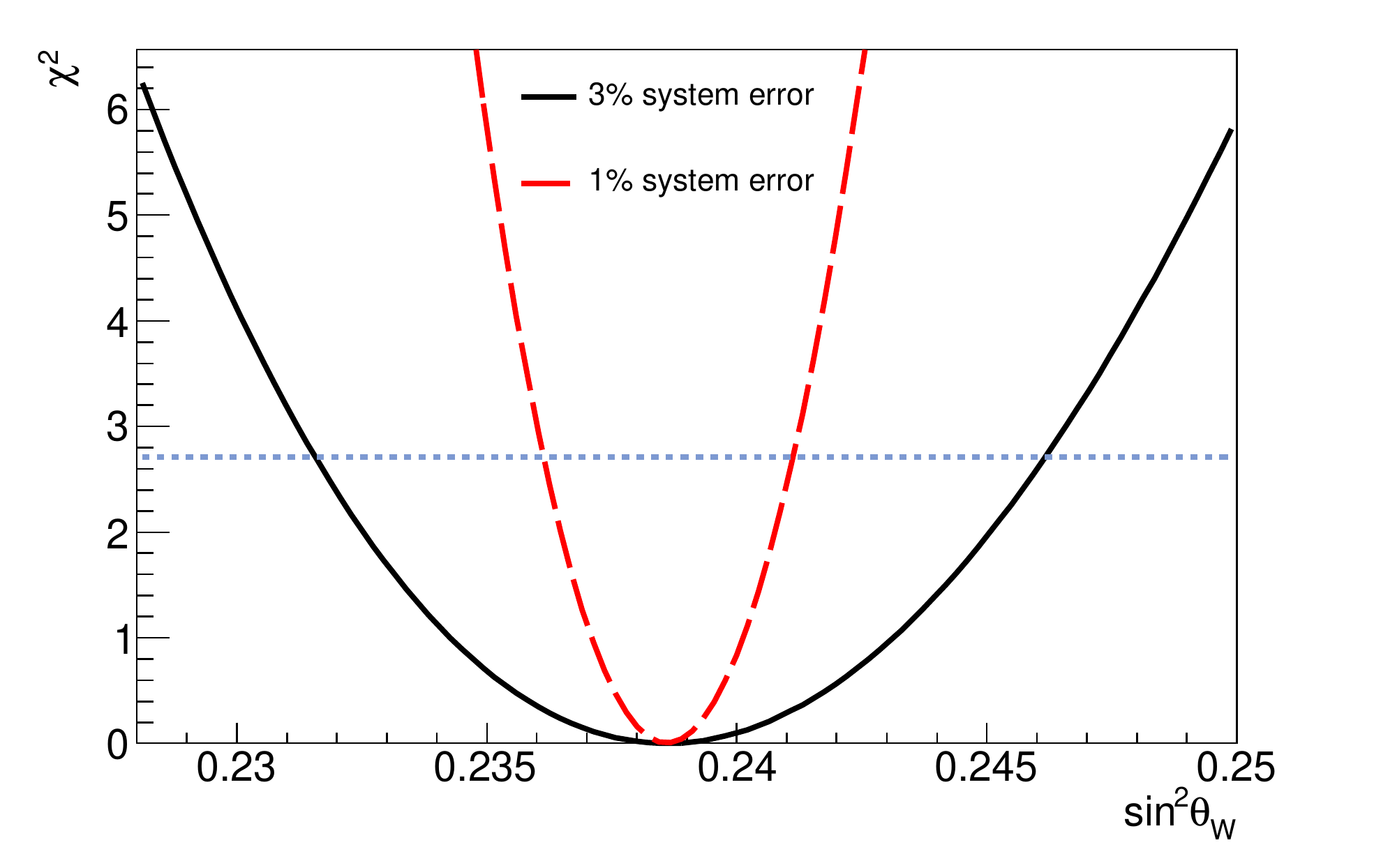}
  \includegraphics[width=0.45\textwidth]{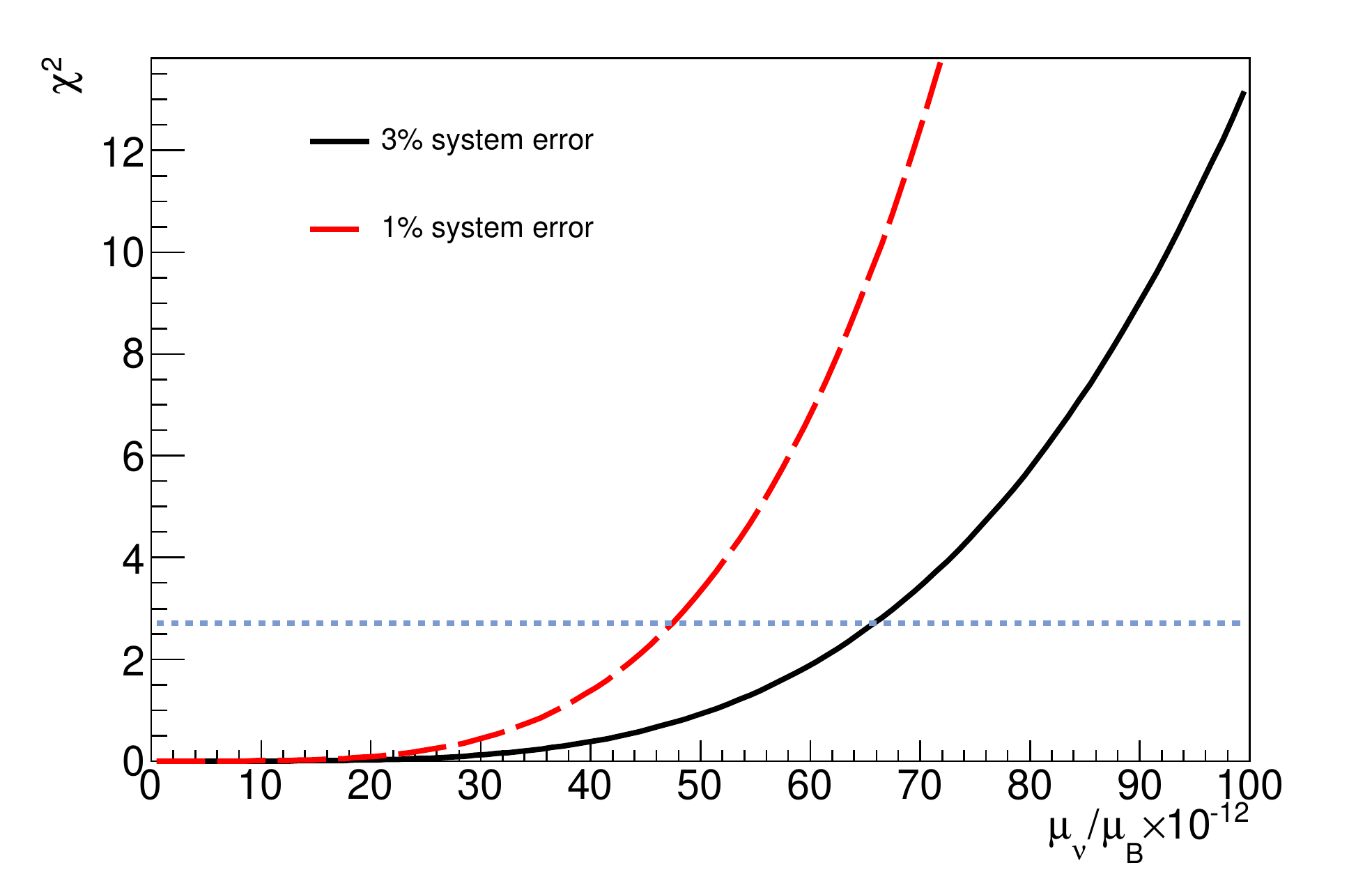}
  \caption{\label{fig:weakanglechis} The expected sensitivity of the weak mixing angle $\sin^2\theta_W$~(left) and of the neutrino magnetic moment $\mu_\nu$~(right).}
  \end{center}
\end{figure}

By minimizing the $\chi^{2}$ function, the expected sensitivities of $\sin^2\theta_W$ and $\mu_{\nu}$ are obtained and shown in Fig.~\ref{fig:weakanglechis}.
The 90\% C.L. of $\sin^2\theta_W$ is 0.232 to 0.246 at keV energy transfer.
The upper limit of the neutrino magnetic moment is ${\rm \mu_{\nu}/\mu_{B}~\textless{}6.6\times 10^{-11}}$.
As a comparison, the best measurement of $\sin^2\theta_W$ at low energy transfer is the atomic parity violation~(APV) experiment, $0.238\pm 0.005$~(1$\sigma$)~\cite{Cadeddu:2019eta}.
From the COHERENT experiment, the value of $\sin^2\theta_W$ is $0.26^{+0.04}_{-0.03} (1\sigma)$~\cite{Cadeddu:2020lky}.
The best limit for $\mu_{\nu}$ is ${\rm \mu_{\nu}~\textless{}2.9\times 10^{-11}\mu_{B}}$ measured by GEMMA~\cite{GEMMA}.

It should be noted that above sensitivities are obtained by fixing one parameter and minimizing the other one.
For example, $\mu_{\nu}$ is fixed to 0 when studying $\sin^2\theta_W$.
Similarly, $\sin^2\theta_W$ is fixed at the nominal value when searching for the upper limit of $\mu_{\nu}$.
In the future, a joint fitting of $\sin^2\theta_W$, $\mu_{\nu}$, and Y$_{e^{-}}$ can be performed based on the large CE$\nu$NS data set.

\section{Summary}
\label{summary}

In this paper, we have put forward a 200~kg dual-phase argon time projection chamber to detect the CE$\nu$NS from the Taishan nuclear reactor.
The outer shielding and the active veto are designed.
The backgrounds from ambient radioactivity and cosmic-ray muons are simulated.
With the help of a single-phase argon detector surrounding the dual-phase TPC, the background level is 153 per day in the energy range of interest.
This compares to a CE$\nu$NS signal rate of 955 per day.
With the large data set, the expected sensitivities of the weak mixing angle $\sin^2\theta_W$, the neutrino magnetic moment $\nu_\mu$ are presented.
In addition, a synergy between the reactor antineutrino CE$\nu$NS experiment and the WIMP dark matter experiment is foreseen, not only in the better understanding of the neutrino floor, but also in the understanding of the detector response at keV nuclear recoil energy range.

\section*{Acknowledgement}
\addcontentsline{toc}{section}{Acknowledgement}

The study is supported by National Key R\&D Program of China (2016YFA0400304) and National Natural Science Foundation of China (11975244).

The authors would like to thank Yufeng Li, Yiyu Zhang, and Yi Wang for the helpful discussion.

\bibliographystyle{unsrt}

\bibliography{CEvNS_Ar_Reactor}

\end{document}